\documentclass{paper}

\newcommand{\la}{\lesssim}
\newcommand{\ga}{\gtrsim}
\newcommand{\real}{\mathbb{R}}

\begin{document}

\title{Projection effects in mass-selected galaxy-cluster samples}
\author{Katrin Reblinsky and Matthias Bartelmann\\
  Max-Planck-Institut f\"ur Astrophysik, P.O.~Box 1523, D--85740
  Garching, Germany}

\date{\em Astronomy \& Astrophysics (1999)}

\begin{abstract}

Selection of galaxy clusters by mass is now possible due to weak
gravitational lensing effects. It is an important question then
whether this type of selection reduces the projection effects
prevalent in optically selected cluster samples. We address this
question using simulated data, from which we construct synthetic
cluster catalogues both with Abell's criterion and an aperture-mass
estimator sensitive to gravitational tidal effects. The
signal-to-noise ratio of the latter allows to some degree to control
the properties of the cluster sample. For the first time, we apply the
cluster-detection algorithm proposed by Schneider to large-scale
structure simulations.

We find that selection of clusters through weak gravitational lensing
is more reliable in terms of completeness and spurious
detections. Choosing the signal-to-noise threshold appropriately, the
completeness can be increased up to 100\%, and the fraction of
spurious detections can significantly be reduced compared to
Abell-selected cluster samples.

We also investigate the accuracy of mass estimates in cluster samples
selected by both luminosity and weak-lensing effects. We find that
mass estimates from gravitational lensing, for which we employ the
$\zeta$-statistics by Kaiser et al., are significantly more accurate
than those obtained from galaxy kinematics via the virial theorem.

\end{abstract}

\maketitle

\section{Introduction}

Observed samples of galaxy clusters are notoriously contaminated by
projection effects (Frenk et al.~1990) even when selected with respect
to their X-ray emission (Bartelmann \& Steinmetz 1996; van Haarlem,
Frenk \& White 1997; Cen 1997). This is particularly annoying because
clusters constitute an important class of cosmological objects, and
cleanly defined cluster samples could provide a wealth of cosmological
information. Not only the definition of clusters within samples, but
also observable cluster properties like their richness, X-ray
luminosity, gas fraction, degree of substructure, velocity dispersion
and the mass estimates derived therefrom can substantially be biased
by projection effects.

While not solving the principal problem of observing aspherical
three-dimensional objects in projection, selection of clusters by mass
rather than luminosity might prove a substantial step towards more
reliably defined cluster samples. With the advent of weak-lensing
techniques and wide-field surveys, mass-selected cluster samples are
beginning to come within reach. The tidal field of the total
gravitating cluster mass determines the amount of distortion observed
in images of background galaxies (Kaiser \& Squires 1993). A variety
of elaborate tools has been developed in the past few years to detect
clusters and derive their masses utilising distortions of faint-galaxy
samples (Fahlman et al.~1994; Kaiser 1995; Kaiser, Squires \&
Broadhurst 1995; Seitz \& Schneider 1995; Bartelmann et al.~1996;
Seitz \& Schneider 1996; Squires \& Kaiser 1996 to name just a
few). In particular, weighted second-order measures of the distortion
field in apertures like the {\em aperture mass\/} statistics ($M_{\rm
ap}$-statistics) provide means to derive the signal-to-noise ratio $S$
of a measurement from the data themselves (Schneider 1996). A possible
strategy to use aperture measures like $M_{\rm ap}$ for cluster
detection is to move the aperture across a wide data field and monitor
$M_{\rm ap}$ and $S$ as a function of aperture position. Points or
regions of high $S$ can then be identified as potential clusters,
leading to the definition of mass-selected cluster samples. This
technique led to a cluster detection in the ESO Imaging Survey
(T.~Erben, private communication). A further spectacular example of
galaxy clusters detected through weak lensing is given by Kaiser et
al.~(1998).

Strictly speaking, distortions caused by gravitational lensing trace
the gravitational tidal field rather than the mass itself. This is due
to the fact that a sheet of constant surface mass density added to any
given lens does not change the shapes of distorted images, but only
their sizes (the so-called mass-sheet degeneracy). Therefore, cluster
detection techniques based on distortion measurements alone select
clusters not by mass, but by the amplitude of their tidal
field. Keeping this in mind, we will hereafter speak of {\em
mass-selected\/} clusters in that sense.

The question then arises, are the cluster samples so obtained any more
reliable than, e.g., cluster samples selected via Abell's criterion or
X-ray luminosity? More precisely, what fraction of true
three-dimensional clusters is detected that way, and what fraction of
the mass-selected samples are spurious detections, either not
corresponding to real clusters at all or to clusters outside the
desired mass range? This is the question addressed in this paper using
simulated $N$-body data, to which we apply for the first time the
$M_{\rm ap}$ statistics to construct synthetic cluster samples.

We use high-resolution large-scale structure simulations and first
identify three-dimensional clusters in real space. We then populate
the simulation volume with galaxies, fixing the average mass-to-light
ratio. In projection, we apply Abell's criterion to construct
optically-selected cluster samples for comparison, and the
$S$-statistics to define mass-selected cluster samples via the
gravitational lensing effects of the simulated matter
distribution. The quality of the mass-selected samples is then
assessed in comparison to that of the optically-selected
samples. Moreover, we derive masses both from galaxy kinematics and
weak lensing to compare the reliability of the different mass
estimates.

Section~2 details the methods used. Section~3 discusses the quality of
the cluster samples in terms of completeness and the fraction of
spurious detections. In Sect.~4 and in the Appendix, we give examples
for the line-of-sight structure of some representative mass-selected
clusters. Mass estimates are presented and discussed in Sect.~5, and
Sect.~6 summarises our conclusions.

\section{Methods}

\subsection{$N$-body simulation}
\label{N_body}

In order to study the influence of projection effects on cluster
surveys selected by optical and gravitational lensing information, we
need simulated data allowing to mimic as accurately as possible the
selection of clusters and the determination of their properties, for
instance their masses. At the same time, the full phase-space
information is required to assess the amount of contamination of
selected clusters by intervening matter along the line-of-sight
(hereafter {\em los\/}). For this purpose, we use a large,
high-resolution $N$-body simulation of a standard Cold Dark Matter
(SCDM) universe. The simulation was carried out within the GIF project
(Kauffmann et al.~1998) using a parallelised version of the HYDRA
code. HYDRA (Couchman, Pearce \& Thomas 1995; Couchman, Thomas \&
Pearce 1996) is an adaptive particle-particle particle-mesh (AP$^3$M)
code. It uses direct force summation in clustered regions, whereas
long-range forces and forces within weakly clustered regions are
computed on a mesh. The simulation used here transports $256^3$
particles in a periodic comoving cubic volume of $(85\,h^{-1}\,{\rm
Mpc})^3$, where $h$ is the Hubble constant in units of $100\,{\rm
km\,s^{-1}\,Mpc^{-1}}$.

The simulation adopts the approximation to the linear CDM power
spectrum (Bond \& Efstathiou 1984) given by
\begin{equation}
  P(k)=\frac{A k}{[1+[aq+(bq)^{3/2}+(cq)^2]^{\nu}]^{2/\nu}}\;,
\end{equation} 
where $q=\Gamma^{-1}\,k$ with the shape parameter $\Gamma$,
$a=6.4\,h^{-1}\,$Mpc, $b=3\,h^{-1}\,$Mpc, $c=1.7\,h^{-1}\,$Mpc, and
$\nu=1.13$. The normalisation constant, $A$, is chosen by fixing
$\sigma_8$, the {\em rms\/} density contrast in spheres of
$8\,h^{-1}\,$Mpc radius. It is determined following the procedure
outlined by White, Efstathiou \& Frenk (1993) to meet the present-day
local cluster abundance of $\approx8\times10^{-6}\, h^3\,{\rm
Mpc}^{-3}$ for rich galaxy clusters. The cosmological parameters are
$H_0=50\,{\rm km\,s^{-1}\,Mpc^{-1}}$ (i.e.~$h=0.5$), $\Omega_0=1$,
$\Omega_\Lambda=0$, $\Gamma=0.5$, and $\sigma_8=0.6$.

We select a simulation box located at a redshift of $z=0.431$ to
achieve high lensing efficiency on sources at redshifts around unity,
where we assume sources to be throughout this paper.

For the analysis of gravitational lensing effects of the simulated
matter distribution, namely the $S$- and $\zeta$-statistics to be
introduced below, the high resolution provided by the GIF simulations
is essential. Spatial and mass resolution must be distinguished. The
spatial resolution is determined by the comoving force softening
length, $l_{\rm soft}=36\,h^{-1}\,$kpc. At one softening length from a
particle, the softened force is about half its Newtonian value. This
limitation is reduced by the high redshift of our simulation box,
where the force softening translates to a very small force softening
angle, $\theta_{\rm soft}\approx10''$. The mass resolution, which
describes the effect of the finite particle number, is given by the
particle mass $m_{\rm p}=1.0\times10^{10}\,h^{-1}\,M_\odot$. The
finite mass resolution introduces a white noise component into the
simulations. This is not negligible for the SCDM model because there a
higher proportion of the particles is in voids than for models with
lower mean density.

Since we want to evaluate two different methods for detecting clusters
{\em in projection\/}, we first have to create a sample of {\em
true\/} 3-dimensional (3-D) clusters or groups from the simulation
that will serve as a reference set.

We extract clusters and groups from the 3-D dark-matter simulation
with a friends-of-friends algorithm (also called group-finder,
cf.~Davis et al.~1985). The friends-of-friends algorithm is based on a
percolation analysis: It identifies groups and clusters in the
simulation box by linking together all particle pairs separated by
less than a fraction $b$ of the mean particle separation. Each
distinct subset of connected particles is then taken as a group or
cluster. We have chosen $b=0.2$, but the result of group finding does
not sensitively depend on the exact choice of $b$.

Assuming that the clusters and groups found by the group-finder are
completely virialised, we compute their virial masses $M_{200}$,
defined as the mass enclosed by a sphere with a radius $r_{200}$ which
contains a mean overdensity of 200 times the critical mass density,
$\bar\rho=200\,\rho_{\rm crit}$. For an $\Omega_0=1$ universe, this
radius approximately separates virialised regions from the infall
regions of the haloes (Cole \& Lacey 1996). Several reference sets of
``true'' 3-D clusters are then formed by selecting objects with
$M_{200}$ above certain mass thresholds.

\subsection{Construction of mock cluster catalogues by optical cluster
  selection}

Having dark-matter particles only, we need to populate our simulation
with galaxies for optical cluster selection. We employ the following
scheme.

Galaxy luminosities $L$ are drawn from a Schechter function (Schechter
1976),
\begin{equation}
  \phi(L) = N_*\,(L/L_*)^{-\alpha}\,\exp{(-L/L_*)}\;,
\end{equation} 
with parameters $L_*=3.77\,\times10^9\,L_\odot$ and $\alpha=1$ taken
from the CfA redshift survey by Marzke, Huchra \& Geller (1994). The
formal divergence for $L\to0$ in the number-density integral of the
luminosity function is avoided by introducing a lower luminosity
cut-off $L_0=0.1\,L_*$.  For the normalisation of the luminosity
distribution, we follow the prescription by Schechter (1976). We
calculate a richness estimate by computing the most probable value of
the third-brightest absolute magnitude $M_3$, and then integrate the
luminosity distribution from $M_3$ to $M_{3}+2$. Frenk et al.~(1990)
showed that this yields the dimension-less normalisation factor
$N_*=60.0$. The normalisation factor determines the amount of luminous
galaxies to be introduced into the simulation. The total mass-to-light
ratio of the 3-D clusters turns out to be
$M/L=300\,h\,M_\odot/L_\odot$ on average, in qualitative agreement
with observations.

Assuming that mass follows light in our model universe, galaxies
inherit positions and velocities from randomly selected dark-matter
particles. In this sense, our constructed galaxy sample is unbiased
both in number density and velocity.

Transforming luminosities to apparent magnitudes for higher redshifts,
we account for the $k$ correction. If the spectral energy distribution
varies with frequency $\nu$ as a power law with exponent $\alpha'$,
the additive $k$ correction is
\begin{equation}
  k = -2.5\,(1+\alpha')\,\log_{10}(1+z)\;.
\end{equation}
We choose $\alpha'=-1.5$ for the spectral index, which sufficiently
well reflects the spectral properties of ordinary galaxies.

Volume-limited cluster catalogues are then obtained after projecting
particle positions onto planes along the three orthogonal axes of the
simulation box. Groups and clusters in projection are identified with
a 2-D version of the friends-of-friends algorithm.

We then apply the optical Abell criterion (Abell 1958) to select
galaxy clusters. This widely used cluster detection and classification
scheme does not depend on redshift $z$. Therefore, it can also be
employed at the high redshift of the simulation even though it has
traditionally been used only for fairly shallow cluster
surveys. Briefly, a cluster is classified as an Abell cluster if
within the Abell radius of $r_{\rm a}=1.5\,h^{-1}\,$Mpc from its
centre, and after subtraction of the mean background, the number count
of galaxies exceeds a certain value $n_{\rm a}$. Counting is
restricted to the apparent magnitude interval $(m_3,m_3+2)$, where
$m_3$ denotes the apparent magnitude of the third-brightest cluster
galaxy. The actual count $n_{\rm a}$ is used to assign Abell richness
classes ${\cal R}$. For ${\cal R}=0$, a cluster has to contain at
least $n_{\rm a}=30$ galaxies, while ${\cal R}=1$ and ${\cal R}=2$
correspond to $n_{\rm a}\ge50$ and $n_{\rm a}\ge80$, respectively.

We also straightforwardly apply Abell's criterion to three-dimensional
clusters in order to assess the influence of projection effects on
richness-class estimates.

For the background subtraction, we follow Frenk et al.~(1990). In
order to estimate the background, i.e.~the contamination by foreground
and background galaxies in the simulation box, we assume that the
number of galaxies contributing to the contamination is proportional
to the volume projected onto the cluster. In our case, we expect 8
background galaxies within a cylinder of volume $V_{\rm proj}= \pi
r_a^2 l_{\rm box}$. Therefore, a cluster with richness class ${\cal
R}=1$ has to encompass at least 58 galaxies in the appropriate
magnitude interval; 8 background galaxies in addition to the 50
genuine cluster members.

Since observed column densities towards galaxy clusters will be
considerably larger than assumed here, and since the conditions in
realistic observations are less controlled than here, projection
effects could even be larger in reality.

\subsection{Detection of dark-matter concentrations through weak
  gravitational lensing}
\label{lensing}
	   
\subsubsection{Basic relations}

We briefly review in this section relations from gravitational lensing
theory important later on. For a derivation cf.~Schneider, Ehlers \&
Falco (1992). Some remarks on the numerical calculation of lensing
properties will also be made.

The dimension-less surface mass density $\kappa$ (also called {\em
convergence\/}) is given by
\begin{equation}
  \kappa(\vec x) = \frac{\Sigma(\vec x)}{\Sigma_{\rm cr}}\;,
\end{equation}
with the critical surface mass density
\begin{equation}
  \Sigma_{\rm cr}=\frac{c^2}{4\pi G}\,
  \frac{D_{\rm s}}{D_{\rm d}D_{\rm ds}}\;.
  \label{sig_cr}
\end{equation}
$D_{\rm s}$, $D_{\rm d}$, and $D_{\rm ds}$ are the angular-diameter
distances from the observer to the sources, from the observer to the
lens, and from the lens to the sources, respectively. The surface mass
density $\kappa(\vec x)$ is related to the effective deflection
potential $\psi(\vec x)$ through the Poisson equation
\begin{equation}
  \nabla^2\psi(\vec x) = 2\,\kappa(\vec x)\;,
\label{eqpos}
\end{equation}
which can be solved for $\psi$,
\begin{equation}
  \psi(\vec x)=\frac{1}{\pi}\,\int_{\real^2}\,{\rm d}^2 x'\,
  	       \kappa(\vec x')\,\ln{(|\vec x-\vec x'|)}\;.
\end{equation}
Boundary conditions have to be specified when solving
eq.~(\ref{eqpos}) numerically. Periodic boundaries are adequate
because of the periodicity of the simulation volume.

For numerically computing $\psi$, the projected particle positions are
interpolated on a grid of $2048^2$ cells to maintain the high
resolution of the $N$-body simulation. The resulting surface mass
density is scaled with the critical surface mass density
(\ref{sig_cr}) to find the convergence $\kappa$. For a numerically
stable and efficient method to convert $\kappa$ to $\psi$, we use a
fast Poisson solver (Swarztrauber 1984). The efficiency of this method
rests on the fast Fourier transform (FFT) leading to an asymptotic
operation count of ${\cal O}(2\,N\,\log N)$. The algorithm
approximates the Laplacian on a grid, transforms to Fourier space,
solves the resulting tri-diagonal system of linear equations, and
back-transforms to real space. In contrast to other approaches, the
approximation is made here by discretising the equations, which can
then be solved exactly by a subsequent discrete FFT.

Having determined the deflection potential $\psi(\vec x)$, the local
properties of the lens, such as the surface mass density $\kappa$ and
the complex shear $\gamma\equiv\gamma_1+{\rm i}\,\gamma_2$, can be
expressed in terms of second derivatives of $\psi(\vec x)$,
\begin{eqnarray}
  \kappa(\vec x)    &=& \frac{1}{2}\,\nabla^2\psi(\vec x)\;,\\
  \gamma_{1}(\vec x)&=& -\frac{1}{2}\,(\psi_{,11} - \psi_{,22})\;,\\
  \gamma_{2}(\vec x)&=& -\psi_{,12}\;,
\end{eqnarray}
where indices $i$ following commas denote partial derivatives with
respect to $x_i$. \footnote{Notice the signs of the shear components:
We follow the sign convention of Schneider \& Seitz (1995).}

\subsubsection{Aperture mass measures}
\label{aperturemass}

Galaxy clusters can be selected solely by their mass using the method
developed by Schneider (1996). For detecting dark-matter
concentrations through image distortions of faint background galaxies,
we define an aperture mass measure $m(\vec x)$ as
\begin{equation}
  m(\vec x) \equiv \int{\rm d}^2y\,
  \kappa(\vec y)\,W(|\vec x-\vec y|)\;,
\end{equation} 
where the integral extends over a circular aperture $|\vec x-\vec
y|\le R$, and $W(y)$ is a continuous weight function depending on the
modulus of $\vec y$ only, vanishing for $y>R$. We assume $W(y)$ to be
a compensated filter function, i.e.
\begin{equation}
  \int_{0}^{R}{\rm d}y\,y\,W(y)=0\;.
\end{equation} 
For such filter functions, the aperture mass measure $m(\vec x)$ can
be expressed in terms of the tangential shear component $\gamma_{\rm
t}$ inside a circle with radius $R$ (Fahlman et al.~1994; Schneider
1996)
\begin{equation}
  m(\vec x)=\int{\rm d}^2y\,Q(|\vec y|)\,
  \gamma_{\rm t}(\vec x;\vec y)\;,
\label{mapstat}
\end{equation}
where $Q$ is related to the filter function $W$ by
\begin{equation}
  Q(x) \equiv \frac{2}{x^2}\,\int_{0}^{x}\,{\rm d}x'\,
              x'\,W(x')-W(x)\;.
\end{equation}
In the above equations, the tangential shear $\gamma_{\rm t}(\vec
x;\vec y)$ at the position $\vec y$ relative to position $\vec x$ is
given by
\begin{equation}
  \gamma_{\rm t}(\vec x;\vec y)=\Re[\gamma(\vec y)\,
  {\rm e}^{-2{\rm i}\phi}]\;,
\end{equation}
where $\phi$ is the position angle of $\vec y-\vec x$.
Equation~(\ref{mapstat}) relates the spatially filtered aperture mass
$m(\vec x)$ to the observable shear field.

\subsubsection{Signal-to-noise ratio}
\label{sn_section}

An estimate for the shear field $\gamma$, and thus for the aperture
mass $m(\vec x)$ via equation (\ref{mapstat}), is provided by the
distortions of images of faint background galaxies. The complex
ellipticity of galaxy images, $\chi$, is commonly defined in terms of
second moments of the surface-brightness tensor (Schneider \& Seitz
1995). For sources with elliptical isophotes with axis ratio $r\le1$,
the modulus of the source ellipticities is given as $|\chi^{({\rm
s})}|=(1-r^2)/(1+r^2)$, and the phase of the $\chi^{({\rm s})}$ is
twice the position angle of the ellipse.

It has been demonstrated (Schramm \& Kaiser 1995; Seitz \& Schneider
1997) that the ellipticity $\chi$ of a galaxy image is an unbiased
estimate of twice the local shear in the case of weak lensing,
$|\gamma|\ll1$. If one assumes that the intrinsic orientations of the
sources are random,
\begin{equation}
  \langle\chi^{({\rm s})}\rangle=0\;,
\end{equation}
with the average taken over an ensemble of sources, then all average
net image ellipticities $\langle\chi\rangle\ne0$ reflect the
gravitational tidal effects of the intervening mass distribution. We
draw the source ellipticities from a Gaussian probability distribution
\begin{equation}
  p_{\rm s}(|\chi^{({\rm s})}|) =
  \frac{1}{\pi\sigma_{\chi}^2
  \left[1-\exp\left(-\sigma_\chi^{-2}\right)\right]}\,
  \exp\left(-\frac{|\chi^{({\rm s})}|^2}{\sigma_\chi^2}\right)\;,
\end{equation}
where the width of the distribution is chosen as $\sigma_\chi=0.3$. We
set the number density of the background sources to $n=35\,{\rm
arcmin}^{-2}$.

The complex image ellipticity $\chi$ can then be calculated in terms
of the source ellipticity $\chi^{({\rm s})}$ and the reduced shear
$g\equiv\gamma\,(1-\kappa)^{-1}$ by the transformation (Schneider \&
Seitz 1995)
\begin{equation}
  \chi=\frac{\chi^{({\rm s})}-2g+g^2\chi^{({\rm s})*}}
  {1+|g|^2-2\,\Re(g\chi^{({\rm s})*})}\;.
\end{equation}

In analogy to the tangential shear component $\gamma_{\rm t}$
occurring in (\ref{mapstat}), a similar quantity for the image
ellipticities can be defined. Consider a galaxy image $i$ at a
position $\vec x_i$ with a complex image ellipticity $\chi_i$. The
tangential ellipticity $\chi_{{\rm t}i}(\vec x)$ of this galaxy
relative to the point $\vec x$ is then given by Schneider (1996)
\begin{equation}
  \chi_{{\rm t}i}(\vec x)=\Re\left(\chi_i\,
  \frac{(X_i - X)^*}{(X_i - X)}\right)\;,
\end{equation}
where $X_i=x_{i1}+{\rm i}\,x_{i2}$ and $X=x_1+{\rm i}\,x_2$ are
complex representations of the vectors $\vec x_i=(x_{i1},x_{i2})$ and
$\vec x=(x_1,x_2)$.

We are now able to estimate the weighted integral (\ref{mapstat}) as a
discrete sum over galaxy images,
\begin{equation}
  m(\vec x)\approx\frac{1}{2n}\,\sum_{i}\chi_{{\rm t}i}(\vec x)\,
  Q(|\vec x_i-\vec x|)\;.
\label{map_1}
\end{equation}
The dispersion $\sigma_{\rm m}(\vec x)$ of the aperture mass estimate
$m(\vec x)$ is found by squaring (\ref{map_1}) and taking the
expectation value, which leads to
\begin{equation}
  \sigma_{\rm m}^{2}(\vec x)=\frac{\sigma_\chi^2}{4\,n^2} 
  \sum_{i}Q^2(|\vec x_i-\vec x|)\;.
\end{equation}
Finally, the {\em signal-to-noise\/} ratio $S$ at position $\vec x$
can be written as
\begin{equation}
  S(\vec x) \equiv \frac{m(\vec x)}{\sigma_{\rm m}(\vec x)}=
  \frac{\sqrt{2}}{\sigma_\chi}\,
  \frac{\sum_i\chi_{{\rm t}i}(\vec x)\,Q(|\vec x_i-\vec x|)}
       {\left[\sum_iQ^2(|\vec x_i-\vec x|)\right]^{1/2}}\;.
\label{eq_sn}
\end{equation}

\subsubsection{The $\zeta$-statistics}
\label{zeta_mass}

So far, the formalism for aperture mass measures and their
signal-to-noise ratios is independent of the choice for the weight
function $W$. Specialising $W$ now, we are led to aperture measures
with different merits. Two principal choices for the filter function
have been suggested in the literature. One leads to the
$\zeta$-statistics proposed by Kaiser (1995) and first applied by
Fahlman et al.~(1994). It gives a lower bound to the average surface
mass density $\kappa$ within a circle inside an annulus by measuring
the distortions of background galaxy images inside the annulus. The
$\zeta$-statistics will be used in Sect.~\ref{zeta} for constraining
the masses of clusters detected through their $S$-statistics. The
piece-wise constant weight function for the $\zeta$-statistics reads
(Schneider 1996)
\begin{equation}
  W_\zeta(x) = \left\{\begin{array}{lll}
    \displaystyle
    \frac{1}{\pi x_1^2} & \hbox{for} & 0\le x<x_1 \\
    \displaystyle
   -\frac{1}{\pi (R^2 - x_1^2)} & \hbox{for} & x_1\le x<R \\
    \displaystyle
    0 & \hbox{for} & R\le x<\infty \\
  		\end{array}\right.\;.
\end{equation}
Inserting this weight function into eq.~(\ref{mapstat}) yields
\begin{equation}
  m_\zeta(\vec x) \equiv \zeta(r_1,r_2) =
  \frac{1}{\pi}\,\frac{r_2^2}{r_2^2-r_1^2}\,\int_{r_1}^{r_2}\,
  \frac{{\rm d}^2r}{r^2}\,\gamma_{\rm t}(\vec r)\;,
\label{eq_zeta}
\end{equation}
where $\vec r$ is the distance vector between the point under
consideration and $\vec x$, and $r_1$ and $r_2>r_1$ are the bounding
radii of an annulus around $\vec x$. It can then be shown that
$\zeta(r_1,r_2)$ is related to the mean convergence
$\bar{\kappa}(r_1,r_2)$ in the annulus by
\begin{equation}
  \zeta(r_1,r_2) = \bar\kappa(r_1)-\bar\kappa(r_1,r_2)\;,
\end{equation}
$\bar\kappa(r_1)$ being the mean convergence in the circle with radius
$r_1$ around $\vec x$. In other words, the $\zeta$-statistics
constrains the average convergence in a circular aperture through the
tangential shear in an annulus surrounding the aperture. Since
$\bar\kappa(r_1,r_2)\ge0$, $\zeta(r_1,r_2)$ provides a lower bound to
the mean surface mass density enclosed by $r_1$.

As mentioned before, it is possible to use the image ellipticities
$\chi$ of the background galaxies as unbiased estimates of twice the
tangential shear, $\chi_{\rm t}\approx2\gamma_{\rm t}$. Therefore, the
integral in (\ref{eq_zeta}) can be approximated as a discrete sum over
galaxy images,
\begin{equation}
  \zeta(r_1,r_2) \approx \frac{r_2^2}{N}\,
  \sum_{i=1}^{N}\frac{\chi_{{\rm t}i}}{r_i^2}\;.
\end{equation}

In this study, we want to obtain a lower bound to the total cluster
masses. For a meaningful application of the $\zeta$-statistics, it is
important to include the complete cluster into the measurement. This
can be achieved following Bartelmann (1995). If we apply the
$\zeta$-statistics to a nested set of annuli with radii $r_i$,
$1<i<n$, then the $\zeta$-statistics for an annulus with $r_i<r_j$
reads
\begin{equation}
  \zeta_{ij}=\bar{\kappa}_i-\bar{\kappa}_{ij}\;.
\label{zeta_system}
\end{equation}
where $\bar{\kappa}_i\equiv\bar{\kappa}(r_i)$ and
$\bar{\kappa}_{ij}\equiv\bar{\kappa}(r_i,r_j)$. On the other hand, the
mass $M_{ij}$ in such an annulus is the product of surface mass
density times the area,
\begin{equation}
\label{mass_rings}
  M_{ij}=A_{ij}\,\bar{\kappa}_{ij}\;,
\end{equation} 
where the area of the annulus is
\begin{equation}
  A_{ij}=\pi\,(r_j^2-r_i^2)\;.
\end{equation}
The crucial point is now that the mass contained within a circle of
radius $R$ is always the sum of the masses contained in annuli with
outer radii $r_j<R$, irrespective of how the area is decomposed into
such rings. Keeping this in mind, eqs.~(\ref{zeta_system}) and
(\ref{mass_rings}) can be combined into a system of $(n-1)$ linear
equations with $n$ unknowns $m_k$, where the $m_k$ denote the masses
in adjacent rings $M_{k-1,k}$.

The fact that there is one equation less than the number of unknowns
reflects the scaling invariance of the surface mass density
$\kappa$. Assuming that the outermost annulus does not contain any
significant convergence, i.e.~$M_{n-1,n}=0$, we finally arrive at the
following set of equations for the masses $M_j$ enclosed by radii
$r_j$:
\begin{equation}
  M_j=(1+\frac{A_{1,j}}{A_1})\,m_1-A_{1,j}\zeta_{1,j}
\end{equation}
where $m_1$ is shorthand for
\begin{equation}
  m_1 \equiv A_1\,\frac{A_{1,n}\zeta_{1,n}-A_{1,n-1}\zeta_{1,n-1}}
  {A_{1,n}- A_{1,n-1}}\;.
\end{equation}

Of course, a lower bound to the total cluster mass could also be
obtained by placing an annulus around the entire cluster and applying
the $\zeta$-statistics to that annulus only rather than to a set of
nested annuli. Our approach has two advantages; first, it yields a
profile of $\kappa$ which allows to assess the location of the outer
cluster boundary, for which we found that $R=1.8 h^{-1}$Mpc is an
appropriate choice. Second, it uses galaxy ellipticity measurements in
all annuli rather than the outermost only, thus reducing the
noise. However, the errors in the $M_j$ are correlated at successive
radii, making an immediate interpretation of the significance at any
given radius less transparent.

\subsubsection{The $S$-statistics}

Since the $\zeta$-statistics is not designed for detecting mass
concentrations, its filter function is not optimised for achieving
high signal-to-noise ratios, leading to high noise levels in a
signal-to-noise map. Schneider (1996) solved this problem by
introducing the smooth, continuous weight function
\begin{equation}
  W_{\rm S}(x) = \left\{\begin{array}{ll}
  \displaystyle
  1 & \\
  \displaystyle
  \quad\hbox{for}\quad 0\le x<\nu_1R & \\
  \displaystyle
  \frac{1}{1-c}\left(
    \frac{\nu_1 R}{\sqrt{(x-\nu_1 R)^2+(\nu_1 R)^2}}-c
  \right) & \\
  \displaystyle
  \quad\hbox{for}\quad \nu_1R\le x< \nu_2R & \\
  \displaystyle
  \frac{b}{R^3}\,(R-x)^2(x-\alpha R) & \\
  \displaystyle
  \quad\hbox{for}\quad \nu_2R\le x\le R & \\
  \end{array}\right.\;.
\end{equation}
In the following, the term $S$-statistics refers to the
signal-to-noise ratio obtained from eq.~(\ref{eq_sn}) using the filter
function $W_{\rm S}(x)$, which guarantees low noise in the
signal-to-noise ratio map. The parameters $\alpha$, $b$, and $c$ are
determined once $\nu_1$ and $\nu_2$ are specified; see Schneider
(1996). We choose $\nu_1=0.05$ and $\nu_2=0.8$ in order to achieve
high signal-to-noise ratios and evaluate the $S$-statistics for an
aperture size of 2~arc minutes.

\section{Completeness of cluster catalogues}
\label{complete}

We are now in a position to investigate completeness and homogeneity
of cluster catalogues constructed with the $S$-statistics as opposed
to the optical Abell criterion. To this end, we create two different
samples of 2-D clusters by applying both methods to simulated data
projected along the $x$-, $y$-, and $z$-axes. We then compare these
2-D clusters with our reference set of 3-D clusters.

To assess the quality of the constructed 2-D catalogues, we use
several reference sets of 3-D clusters with different mass ranges.
Looking at different mass ranges instead of cluster richness estimates
is motivated by two reasons. First, the physical quantities of primary
interest are the masses. Furthermore, this kind of comparison is more
suitable for the $S$-statistics, which does not depend on the
distribution of luminous galaxies but of the dark matter only. This
way of addressing projection effects in cluster catalogues differs
from previous studies (e.g.~Cen 1997; van Haarlem et al.~1997), which
focused on the influence of projection effects on the richness
estimate of Abell catalogues.

\begin{table*}
\centering
\caption{\label{astatistic} Completeness and homogeneity of optically
  selected, synthetic cluster catalogues constructed by Abell's
  criterion. For five mass ranges of 3-D clusters, the detection rate
  (det.) and the spurious detection rate (spur.~det.) is shown for
  cluster richness ${\cal R}>0$ and ${\cal R}\ge0$. The detection rate
  gives the percentage of 3-D clusters correctly identified by Abell's
  criterion. The spurious detection rate is the fraction of detected
  clusters which do not correspond to true 3-D clusters within the
  considered mass range. Further explanation is given in the text.}
\begin{tabular}{@{}ccccc@{}}
  3-D clusters &
  \multicolumn{2}{c}{${\cal R}>0$} &
  \multicolumn{2}{c}{${\cal R}\ge0$} \\
  Mass range in $10^{14}\,h^{-1}\,M_\odot$ &
  det. & spur. det. &
  det. & spur. det. \\ [10pt]
  1.32 -- 3.50 & 53\% & 50\% & 66\% & 82\% \\
  1.03 -- 3.50 & 27\% & 50\% & 56\% & 69\% \\
  0.82 -- 3.50 & 17\% & 50\% & 44\% & 64\% \\
  0.55 -- 3.50 & 13\% & 25\% & 36\% & 40\% \\
  0.10 -- 3.50 & 13\% &  6\% & 32\% & 29\% \\
\end{tabular}
\end{table*}

\begin{figure}[ht]
  \includegraphics[width=\columnwidth]{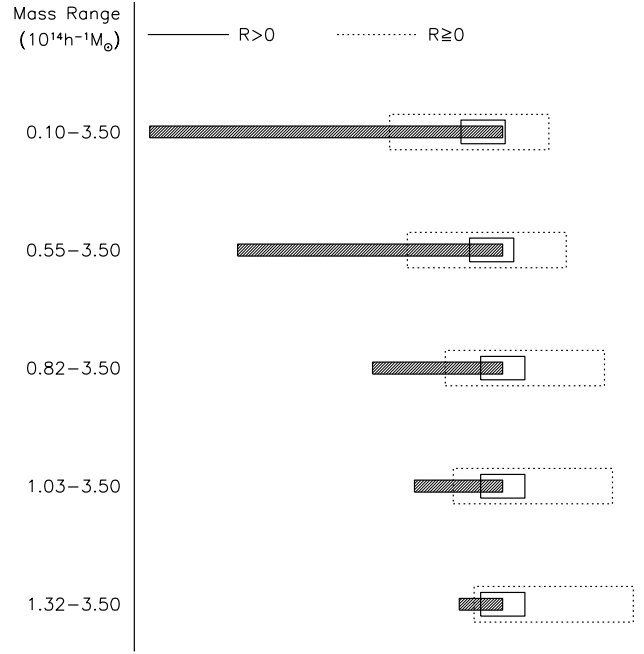}
\caption{Performance of Abell's criterion in identifying clusters in
  different mass ranges (cf.~Tab.~\ref{astatistic}). The hatched bars
  illustrate the number of 3-D clusters for each mass range, the solid
  and dotted bars the number of Abell clusters with ${\cal R}>0$ and
  ${\cal R}\ge0$, respectively. The overlap between hatched and empty
  bars illustrates the fraction of correct detections, and the
  fraction of spurious detections by Abell's criterion corresponds to
  the rest of the empty bars. The fraction of correctly identified 3-D
  clusters and the fraction of spurious detections both increase for
  more massive cluster samples.}
\label{fig1}
\end{figure}

The results for the Abell-selected cluster catalogues are summarised
in Tab.~\ref{astatistic} and Fig.~\ref{fig1}. The first column of
Tab.~\ref{astatistic} lists the mass range of the 3-D cluster
reference set. The next two columns show the fraction of 3-D clusters
correctly detected by Abell's criterion, and the fraction of 2-D
objects which do not correspond to 3-D clusters within the chosen mass
range, respectively.  The fraction of detected clusters is given with
respect to the 3-D clusters in the given mass range, while the
spurious detections are given relative to the total number of 2-D
detections. The last two columns show the same information for a
larger sample of Abell clusters also including clusters of richness
class ${\cal R}=0$. Figure~\ref{fig1} illustrates the information of
Tab.~\ref{astatistic} as a histogram.

The first mass range considered in Tab.~\ref{astatistic},
$(1.32-3.5)\times10^{14}\,h^{-1}\,M_\odot$, reflects the masses
expected for Abell clusters with richness class ${\cal R}>0$. Looking
at absolute numbers, we find that the total number of 2-D clusters
with ${\cal R}>0$ in the Abell catalogue is very similar to the number
of 3-D objects in this mass range (in Fig.~\ref{fig1}, the
corresponding bars are comparably long). However, as
Tab.~\ref{astatistic} shows, only $53\%$ of the 3-D clusters from the
reference set can be found in the 2-D sample of ${\cal R}>0$
clusters. On the other hand, a high percentage ($50\%$) of the 2-D
Abell clusters does not correspond to a true 3-D object. This means
that not only half of the 3-D clusters in this mass range are missed
by Abell's criterion, but also a lot of spurious 2-D objects are
found. This is due to two competing effects occurring in
projection. Intrinsically rich clusters may disappear in the
background, while the richness class of poor clusters can be enhanced
by small groups and field galaxies collected along the line of
sight. In the above case these two effects approximately cancel, so
that the total numbers are approximately correct.

These results are consistent with the findings of Frenk et al.~(1990)
for projection effects in CDM-like universes with different biasing
parameters $b$. Comparing Tab.~2 in their paper with our results for
Abell ${\cal R}=1$ clusters, we find similar projection effects for
our model universe and their CDM-like universes with biasing
parameters between $b=2.5$ and $b=1.6$. A direct comparison is
difficult because of the different normalisations of their model
universe and ours. Furthermore, the redshift dependence of the biasing
parameter $b$ is not known, further complicating a detailed
comparison.

For 2-D galaxy clusters with richness class ${\cal R}=0$, the fraction
of detected 3-D clusters is slightly increased from 53\% to 66\%. At
the same time, the number of spurious 2-D clusters, i.e.~clusters
which cannot be linked to 3-D objects in the reference set, is
increased by more than $30\%$. A detailed analysis of the
line-of-sight structure of these clusters reveals that this large
number of spurious detections is partly due to the additive projection
of poorer groups corresponding to lower-mass 3-D objects. This
increase in the number of both detections and spurious detections
reflects the enhancement or reduction of cluster richness classes due
to projection.

Extending the reference set of 3-D clusters to lower masses
substantiates this assumption. The number of spurious detections
declines quite steeply from over $80\%$ for the mass range of
$(1.32-3.5)\times 10^{14}\,h^{-1}\,M_\odot$ to below $30\%$ for a
lower mass threshold of $0.1\times 10^{14}\,h^{-1}\,M_\odot$, which is
nearly one order of magnitude smaller than the lower mass threshold
for ${\cal R}>0$ clusters. Therefore, many of the 2-D clusters
detected by Abell's criterion do indeed correspond to true 3-D
objects, but in very different mass ranges. This clearly indicates
that for our model universe a change of the richness estimate due to
projection is likely. But still the number of truly spurious
detections, i.e.~detections of 2-D objects which cannot be connected
with any 3-D object, remains quite high even in the broadest mass
range.

\begin{figure}[ht]
  \includegraphics[width=\columnwidth]{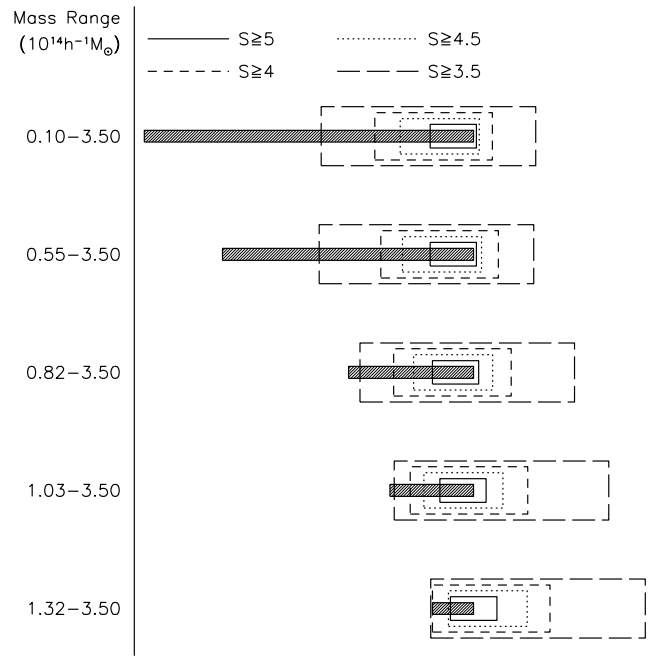}
\caption{Performance of the $S$-statistics in identifying clusters in
  different mass ranges (cf.~Tab.~\ref{sstatistic}). See the caption
  of Fig.~\ref{fig1} for the meaning of the overlapping and
  non-overlapping parts of the bars. The different line types are for
  different $S$ thresholds, as indicated in the plot. With increasing
  $S$, the fraction of spurious detections and the completeness are
  both reduced.}
\label{fig2}
\end{figure}

Turning to the performance of the $S$-statistics in constructing a
complete and homogeneous catalogue, Tab.~\ref{sstatistic} and
Fig.~\ref{fig2} display the results of the $S$-statistics in a manner
analogous to Tab.~\ref{astatistic} and Fig.~\ref{fig1} for Abell's
criterion. Again, the first column contains the mass range of the
investigated 3-D reference set, while the following columns display
the percentage of detected 3-D clusters and of spuriously detected 2-D
objects above a certain $S$-value. The analysis is performed for
objects detected above different signal-to-noise thresholds.

\begin{table*}
\centering
\caption{\label{sstatistic} Completeness and homogeneity of a
  catalogue constructed with the $S$-statistics based on a weak
  gravitational lensing analysis of distorted images of background
  sources. The percentage of detections (det.) and spurious detections
  (spur.~det.) is given for different $S$-values. For further
  explanation, see Tab.~\ref{astatistic} and the text.}
\begin{tabular}{@{}ccccccccccc@{}}
  3-D clusters &
  \multicolumn{2}{c}{$S \ge 5$} &
  \multicolumn{2}{c}{$S \ge 4.5$} &
  \multicolumn{2}{c}{$S \ge 4$} &
  \multicolumn{2}{c}{$S \ge 3.5$} &
  \multicolumn{2}{c}{$S \ge 3$} \\
  Mass range in $10^{14}\,h^{-1}\,M_\odot$ & 
  det. & spur. det. &
  det. & spur. det. &
  det. & spur. det. &
  det. & spur. det. &
  det. & spur. det. \\ [10pt]
  1.32 -- 3.50 & 60\% & 50\% & 60\% & 68\% &100\% & 65\% &100\% & 80\%
  &      &      \\
  1.03 -- 3.50 & 43\% & 27\% & 60\% & 37\% & 76\% & 46\% & 93\% & 63\%
  &      &      \\
  0.82 -- 3.50 & 33\% & 11\% & 48\% & 24\% & 64\% & 32\% & 91\% & 47\%
  &      &      \\
  0.55 -- 3.50 & 18\% &  6\% & 28\% & 10\% & 37\% & 21\% & 61\% & 28\%
  & 81\% & 32\% \\
  0.10 -- 3.50 & 14\% &  6\% & 22\% &  7\% & 30\% & 16\% & 46\% & 29\%
  & 90\% & 22\% \\
\end{tabular}
\end{table*}

The first $S$-threshold investigated in detail is $S\ge5$. This value
has been advocated in the literature, e.g.~by Schneider \& Kneib
(1998), as a signal-to-noise ratio promising significant
detections. In comparison to the optical Abell criterion, the
$S$-statistics has a similar detection rate for Abell ${\cal R}>0$
like objects ($53\%$ with Abell's criterion compared to $60\%$ with
the $S$-statistics). The number of spuriously detected objects
($50\%$) is identical to that for Abell's criterion in the highest
mass range.

\begin{figure}[ht]
  \includegraphics[width=\columnwidth]{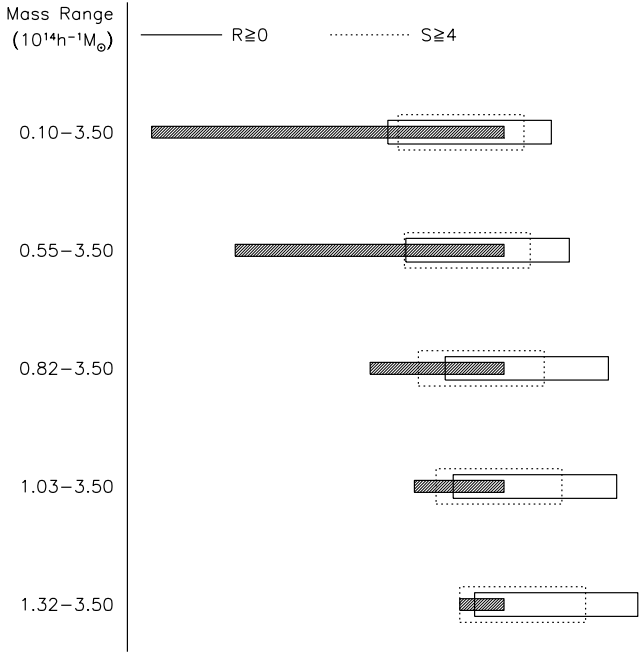}
\caption{Comparison of the performances of the $S$-statistics and
  Abell's criterion in identifying clusters in different mass
  ranges. See the caption of Fig.~\ref{fig1} for the meaning of the
  overlapping and non-overlapping parts of the bars. The solid and
  dotted bars are for Abell-clusters with ${\cal R}\ge0$ and
  $S$-selected clusters with $S\ge4$, respectively. Evidently, the
  $S$-statistics performs better in all mass ranges in terms of
  spurious detections, and completeness is generally also larger.}
\label{fig3}
\end{figure}

The differences between the two methods show up when detections and
spurious detections at mass ranges with a lower mass threshold are
considered. Looking at the detection rate of spurious 2-D objects, we
see a much steeper decline as in the Abell case. For a lower mass
threshold of $1.03\times 10^{14}\,h^{-1}\,M_\odot$, only $27\%$ of the
$S$-detected clusters do not correspond to 3-D clusters of the
reference set, whereas more than half of the Abell clusters in that
mass range are spurious detections. For an even lower mass threshold
of $0.82\times 10^{14}\,h^{-1}\,M_\odot$, the rate of spurious
detections falls to only $11\%$, which clearly indicates that a large
number of suspected spurious detections in reality corresponds to
3-dimensional matter concentrations of lower mass.

If we reduce the $S$-threshold for significant detections to
e.g.~$S\ge4.5$ or even below, the detection rate of 3-D clusters
increases, which means that the $S$-cluster catalogue becomes more
complete. However, the trade-off for the completeness is a higher
number of spurious detections, belonging to lower-mass 3-D mass
concentrations. For an $S$-threshold of $4.0$, we are able to
construct a catalogue containing {\em all\/} massive Abell-like
clusters at the expense of also detecting many less massive 3-D
objects. Figure~\ref{fig3} compares the performance of Abell's
criterion with ${\cal R}\ge0$ and the $S$-statistics with $S\ge4$. It
shows that $S$-selected cluster samples contain fewer spurious
detections and are generally more complete than Abell-selected
samples.

\begin{figure}[ht]
  \includegraphics[width=\columnwidth]{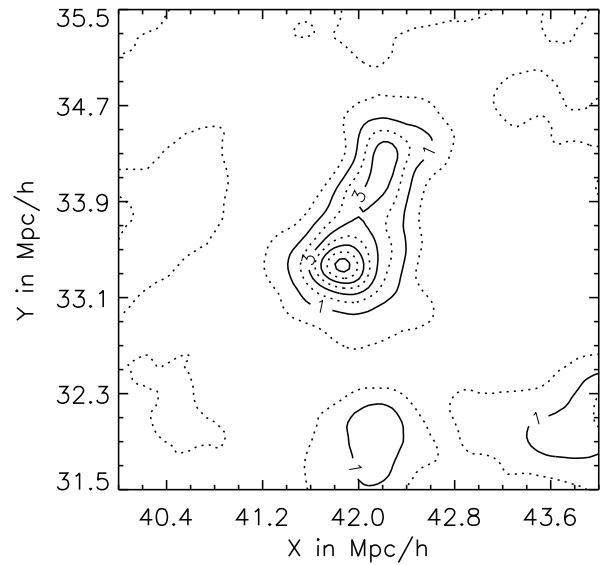}
\caption{\label{dpeak} $S$-map for the rectangular section in
  Fig.~\ref{cms_1} showing the double-peak structure of the map in
  more detail. Contours are spaced by $\Delta S=1$. The structure
  features two maxima, but corresponds to a single 3-D cluster.}
\end{figure}

\begin{figure}[ht]
  \includegraphics[width=\columnwidth]{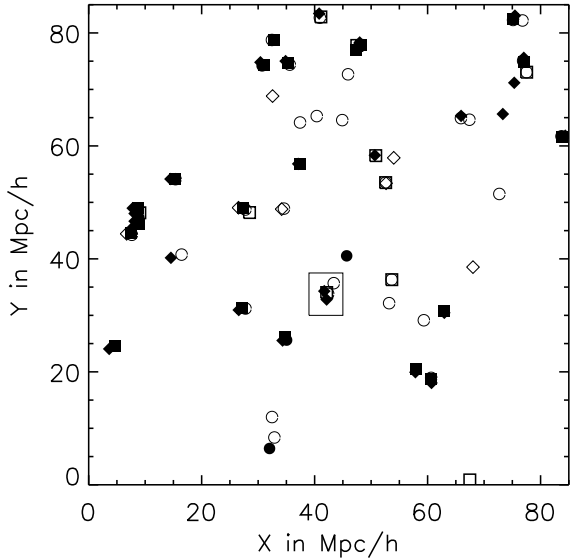}
\caption{\label{cms_1} Projection of true 3-D clusters/groups
  (squares) along the $z$-axis onto the full $xy$-plane of the
  simulation box, and comparison with Abell clusters identified in
  projection (2-D-Abell clusters, circles), and mass-selected clusters
  identified with the $S$-statistics (diamonds). 3-D clusters with
  masses in the range of $(0.10$-$3.5)\times
  10^{14}\,h^{-1}\,M_\odot$, 2-D clusters with ${\cal R}\ge 0$
  (Abell), and $S \ge 3.0$ ($M_{\rm ap}$) are shown. Filled symbols
  refer to clusters with masses $\ge
  0.55\times10^{14}\,h^{-1}\,M_\odot$, ${\cal R}>0$ and $S\ge 3.5$.
  (The offset between close symbols is deliberate to facilitate
  reading; it does not reflect the accuracy of measuring cluster
  positions.)}
\end{figure}

A crucial point in the application of the $S$-statistics is the
identification of peaks in the $S$-map. Following Schneider (1996), we
use a circular aperture for the $S$-statistics, which leads to an
increased sensitivity for round objects. However, some of the $S$-maps
for our simulation data show extended, non-circular areas with
significant $S$-signals. Several of these structures contain more than
one peak coming from within a plateau of high $S$ (see
Fig.~\ref{dpeak}). It is important to properly categorise these
structures as belonging to a {\em single\/} 3-D object and to not
count them twice. An example of such a situation is shown in
Fig.~\ref{dpeak}. The contour plot shows a blow-up of the $S$-map
around two peaks which almost overlap in the lower-resolution $S$-map
of the whole simulation box (see the mark in Fig.~\ref{cms_1}). These
two peaks in the $S$-map correspond to one of the most massive
clusters in the simulation. This is reflected in the high $S=7.6$ of
the higher peak, while the second peak has $S=3.6$. On the other hand,
the sample also contains examples of 3-D clusters projected onto each
other showing only a single featureless peak. Therefore, we conclude
that the morphological information contained in the $S$-map is
low. Both cases of $S$-signals will be discussed in more detail in
Sect.~\ref{structure}.

Summarising the quantitative results from both cluster detection
methods, we can say that the $S$-statistics leads to better results
than Abell's criterion. The catalogues constructed with the
$S$-statistics are more complete and suffer less from spurious
detections, at least in the sense that most peaks correspond to true
3-D objects. The $S$-statistics evidently produces fewer truly
spurious detections than Abell's criterion. However, we note that it
is not possible to obtain a complete catalogue by counting only peaks
with a high signal-to-noise value $S\ge5$. There is no strict
correlation between the height of a signal in the $S$-statistics and
the mass of identified 3-dimensional objects.

\section{Structure of representative clusters}
\label{structure}

To achieve a deeper understanding of projection effects, we study the
structure of archetypical clusters or groups along the line-of-sight.
The main emphasis in this section will be on clusters selected by the
$S$-statistics. A detailed discussion of the structure of Abell
selected clusters along the line-of-sight for both real space and
velocity space can be found, e.g., in Cen (1997) and van Haarlem et
al.~(1997).

Even though this section will concentrate on clusters detected by the
$S$-statistics, in some cases also Abell-selected clusters will be
discussed if the $S$-selected clusters also satisfy Abell's
criterion. As Fig.~\ref{cms_1} shows for one of the three projection
directions, this is the case for a lot of $S$-selected objects,
i.e.~there is considerable overlap between the two selection
methods. Abell's criterion detects the visible light from galaxies,
while the $S$-statistics is sensitive to the underlying distribution
of dark matter, making it possible to construct a ``mass-selected''
sample of clusters, as opposed to ``flux-limited'' samples which are
obtained by observing luminous galaxies. Since this study is performed
under the supposition that both selection methods detect the same
physical objects, we have to assume that luminous galaxies are good
tracers of the dark matter distribution. This is secured by the
assumption used to populate the simulation that light follows mass.

For a more detailed analysis, the $S$-selected clusters will be
subdivided into three classes according to the $S$-threshold
employed. The first class considered are clusters detected with
$S\ge5$, the next class contains clusters with $5>S\ge4$, and the last
class clusters or groups with $4>S\ge3$. The division into classes
according to signal-to-noise values allows an investigation of
systematic differences of projection effects in the different classes.

It is expected from theory that higher-mass clusters lead to larger
values in the $S$-map. Such a trend was found in Sect.~\ref{complete},
but there is no sharp correlation between the masses of detected 3-D
clusters and the threshold imposed on the $S$-statistics. This can be
explained in terms of the intervening matter along the
line-of-sight. Lower-mass 3-D objects are more prone to projection
effects in the sense that the intervening matter makes up a more
substantial fraction of their mass.  Therefore, projection effects
become more important for clusters with lower $S$.  In the case of
clusters with intrinsically lower masses, less intervening matter is
needed to alter the signal in the $S$-map.

\subsection{$S$-statistics: $S\ge5$}

As discussed in Sect.~\ref{complete}, there is a good correspondence
between $S\ge5$ clusters and massive 3-D-clusters. Investigating the
line-of-sight structure of these 2-D-clusters, we can generally state
that nearly all of them show a high, pronounced peak in the position
histogram at the position of the true 3-D cluster. Even though the
amount of contamination with intervening matter in this group is only
moderate, some velocity histograms deviate significantly from a
Gaussian shape.

\begin{figure}[ht]
  \includegraphics[width=\columnwidth]{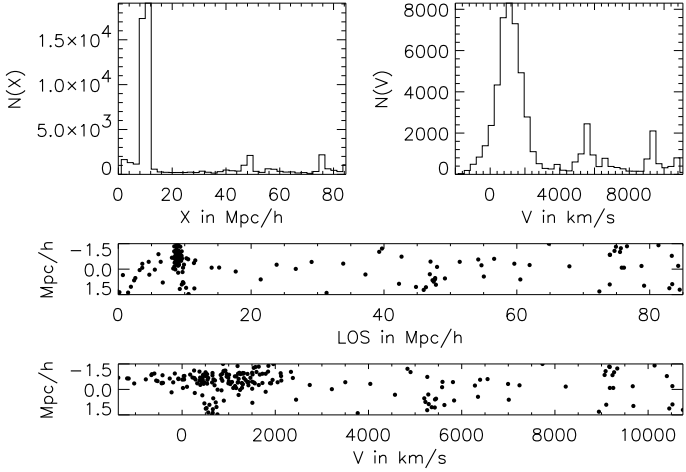}
\caption{\label{cluster_8x} Structure of a large 3-D cluster with
  moderate contamination, detected as a double peak with $S=7.3$ and
  $S=3.6$ (cf.~Fig.~\ref{dpeak}). It is also an Abell ${\cal R}=1$
  cluster, both in projection and in 3-D. The upper left panel
  displays the histogram of the dark matter distribution in real space
  along the line-of-sight, and the upper right panel shows the
  corresponding velocity histogram. The two lower panels show the dark
  matter distribution along a cylinder of radius $1.5\,h^{-1}\,$Mpc in
  real space (along the line-of-sight) and in velocity space. For
  better display in both lower panels, a random fraction of the dark
  matter particles is omitted.}
\end{figure}

The cluster given in Sect.~\ref{complete} as an example for a cluster
exhibiting a double-peaked $S$-map (see Fig.~\ref{dpeak}) clearly
belongs into this class, since the main peak has $S=7.3$, while the
nearby second peak has only $S=3.6$. The structure along the
line-of-sight in both real space and velocity space is shown in
Fig.~\ref{cluster_8x}. The position histogram is characterised by a
dominant peak at the position of the corresponding 3-D cluster which
has a mass of $M_{\rm 3-D}=3.4\times 10^{14}\,h^{-1}\,M_\odot$.  The
amount of dark matter along the line-of-sight is moderate with only
two small clumps $40\,h^{-1}\,$Mpc and $65\,h^{-1}\,$Mpc behind the
main clump, which both have masses smaller than
$10^{13}\,h^{-1}\,M_\odot$.  Although the 3-D cluster has Abell
richness class ${\cal R}=0$, the projected cluster satisfies the 2-D
Abell criterion for ${\cal R}=1$, indicating an inflation of richness
class. The velocity dispersion $\sigma_{\rm 3-D}=775\,{\rm
km\,s^{-1}}$ of the 3-D cluster is reduced to $\sigma_{2-D}=676\,{\rm
km\,s^{-1}}$ in projection, hinting at an asymmetric velocity
ellipsoid of the 3-D cluster.

Although the cluster is only moderately affected by projection
effects, the {\em los\/} velocity histogram strongly deviates from a
Gaussian, which is also true for the velocity distribution of the
3-D-cluster alone. The deviation from Gaussianity, as measured by
higher order moments of the distribution like the skewness ${\cal S}$
and the curtosis ${\cal K}$, is ${\cal S}_{\rm 3-D}=-0.62$ and ${\cal
K}_{\rm 3-D}=0.47$ for the 3-D-cluster as opposed to ${\cal S}_{\rm
2-D}=1.57$ and ${\cal K}_{\rm 2-D}=3.31$ in projection. The
substantial increase of skewness and curtosis in projection emphasises
the influence of projection effects in velocity space. Together with
the decrease of the velocity dispersion and the increase of richness
class in projection, this hints at the presence of non-virialised
sub-clumps in the vicinity of the main cluster. Yet, this detection
corresponds to a very massive 3-D cluster. More examples are given in
Figs.~\ref{cluster_25x} and \ref{cluster_26y} in the Appendix.

\subsection{$S$-statistics: $5>S\ge4$}

The next sample of $S$-detected clusters has lower signal-to-noise;
but, as has been shown in Sect.~\ref{complete}, the mass-selected
cluster catalogue becomes more complete if these clusters are
included. Most of them correspond to true 3-D clusters. Even though
these detections are significant and contain some massive clusters,
their amount of contamination in relation to their main 3-D cluster is
generally larger than for clusters detected with larger $S$. One
typical example of this class is shown in Fig.~\ref{cluster_13x}, two
more are given in Figs.~\ref{cluster_14x} and \ref{cluster_6z} in the
Appendix.

\begin{figure}[ht]
  \includegraphics[width=\columnwidth]{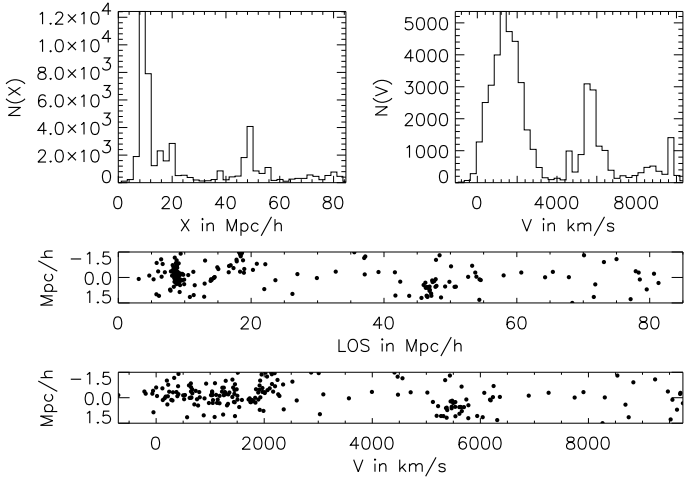}
\caption{\label{cluster_13x} Structure of a moderately contaminated
  3-D cluster with $S=4.7$. See the caption of Fig.~\ref{cluster_8x}
  for a description of the panels.}
\end{figure}

The cluster in Fig.~\ref{cluster_13x} is a high-mass cluster with
$M_{\rm 3-D}=1.5\times 10^{14}\,h^{-1}\,M_\odot$, a clump nearby, and
a second mass clump $30\,h^{-1}\,$Mpc away. It is detected at
$S=4.7$. The velocity dispersion of the projected cluster is broadened
from $\sigma_{\rm 3-D}=697\,{\rm km\,s^{-1}}$ to $\sigma_{\rm
2-D}=865\,{\rm km\,s^{-1}}$, and the projected velocity distribution
has a bimodal shape. The higher-order moments of the velocity
distribution indicate this through skewness and curtosis in projection
(${\cal S}_{\rm 2-D}=1.85$ and ${\cal K}_{\rm 2-D}=3.27$), compared to
3-D (${\cal S}_{\rm 3-D}=0.18$ and ${\cal K}_{\rm 3-D}=-0.40$). The
cluster satisfies Abell's criterion with richness class ${\cal R}=0$,
both in projection and in 3-D.

\subsection{$S$-statistics: $4>S\ge3$}

\begin{figure}[ht]
  \includegraphics[width=\columnwidth]{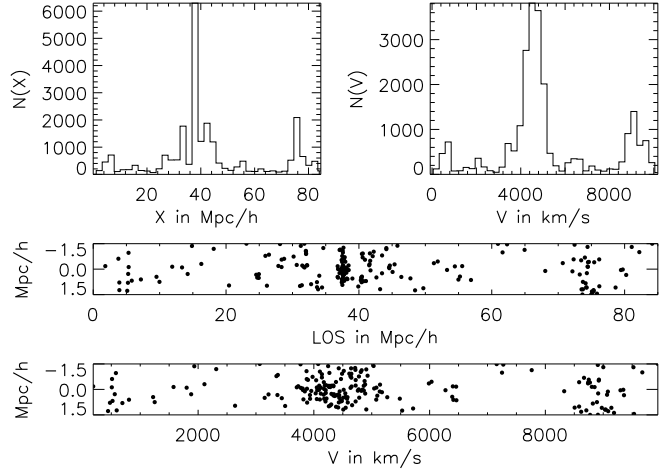}
\caption{\label{cluster_38x} Line-of-sight structure of a highly
  contaminated cluster. This cluster is detected at $S=3.2$. See the
  caption of Fig.~\ref{cluster_8x} for a description of the panels.}
\end{figure}

The last class considered contains clusters with $S$ between 3 and
4. This class is mainly discussed for reasons of
completeness. Clusters identified with such signal-to-noise correspond
to lower-mass objects making projection effects along the
line-of-sight more important. This can be seen looking at the three
examples in Fig.~\ref{cluster_38x} and in Figs.~\ref{cluster_12z} and
\ref{cluster_39x}.

The peak with $S=3.2$ corresponding to the cluster shown in
Fig.~\ref{cluster_38x} is due to a 3-D cluster with a rather low mass
only, $M_{\rm 3-D}=5.9\times 10^{13}\,h^{-1}\,M_\odot$. Compared to
the low 3-D cluster mass, there is a high amount of intervening matter
with several mass clumps along the line-of-sight. Because of this
substantial contamination, the projected velocity dispersion is
overestimated; $\sigma_{\rm 2-D}=669\,{\rm km\,s^{-1}}$ compared to
$\sigma_{\rm 3-D}=364\,{\rm km\,s^{-1}}$ for the 3-D cluster. The
velocity histogram has a second peak at the high-velocity tail of the
distribution. The higher-order moments of the velocity distribution of
the 3-D cluster (${\cal S}_{\rm 3-D}=-0.05$ and ${\cal K}_{\rm
3-D}=-0.07$) change by a large amount when looking at the projected
velocity distribution (${\cal S}_{\rm 2-D}=1.12$ and ${\cal K}_{\rm
2-D}=2.43$). This reflects the large influence the intervening matter
exerts on observation.

\section{Mass estimates}
\label{zeta}

The previous two sections put emphasis on the completeness of
catalogues constructed with two different selection criteria, Abell's
criterion and the $S$-statistics based on weak gravitational
lensing. Furthermore, we investigated the structure of some detected
clusters along the line-of-sight. Of course, both kinds of information
are important when deriving statistical information from such
catalogues. A third very important test for cosmological theories are
the different mass estimates and their relationship with each other. A
mass estimate closely related to the optical selection derives from
the virial theorem. As a gravitational-lensing based mass estimate, we
choose the $\zeta$-statistics, which is closely related to the
$S$-statistics as demonstrated in Sect.~\ref{zeta_mass}.

Under the assumption that clusters of galaxies are bound
self-gravitating systems in dynamical equilibrium, the total cluster
mass can be estimated via the virial theorem (Binney \& Tremaine 1987;
Sarazin 1986),
\begin{equation}
  M_{\rm tot}=\frac{R_{\rm G}\langle v^2\rangle}{G}\;,
\end{equation}
where $R_{\rm G}$ is the gravitational radius of the cluster relating
the system's mass to its potential energy. Gunn \& Gott (1972) showed
that this radius is approximately given by $r_{200}$, the radius of a
sphere containing an overdensity of $200\,\rho_{\rm crit}$. For the
clusters of our study, a radius of $R_{\rm G}=0.75\,h^{-1}\,$Mpc is a
good approximation. Observationally, only the {\em los\/} velocity
dispersion $\sigma_\parallel$ can be measured. Assuming isotropic
orbits, the two quantities are related by $\langle
v^2\rangle=3\sigma_\parallel^2$.

When calculating the radial velocities from simulated data, the Hubble
flow has to be added to the peculiar velocities of the
simulation. Since the simulation data are at high redshift, the
dependence on redshift of the cosmological constants also has to be
taken into account. Therefore, the radial velocities are given by
\begin{equation}
  v_\parallel=\dot{a}(z)\,x_\parallel+a(z)\,\dot x_\parallel =
  a(t)\,[H(z)\,x_\parallel+\dot x_\parallel]\;,
\end{equation}
with the expansion factor
\begin{equation}
  a(z)=(1+z)^{-1}
\end{equation}
and the Hubble parameter
\begin{equation}
  H(z)=H_0\,(1+z)^{3/2}\;.
\end{equation}

Apart from the validity of the virial theorem, the virial mass
estimate depends solely on a correct estimate of the velocity
dispersion $\sigma_\parallel$. Since the velocity dispersion is very
sensitive to field galaxies and small sub-clumps projected onto the
main clusters, it is important to remove these from the sample. We
convolve the velocity histogram with a $4000\,{\rm km\,s^{-1}}$ wide
top-hat filter to reject interlopers, i.e.~galaxies with relative
velocities greater than $4000\,{\rm km\,s^{-1}}$ from the peak of the
convolved histogram are removed. We further employ the so-called
$3\sigma$-clipping procedure proposed by Yahil \& Vidal (1977) which
has widely been applied to observational samples. It can be summarised
as follows:

\begin{enumerate}
\item compute the mean radial velocity;
\item remove the galaxy which deviates most from the mean of the
  sample and re-determine the mean without this galaxy;
\item if the removed galaxy deviates from the new mean by more than
 $3\sigma$, it is removed from the sample;
\item repeat the procedure until the last tested galaxy remains in the
  sample.
\end{enumerate}

\begin{figure}[ht]
  \includegraphics[width=\columnwidth]{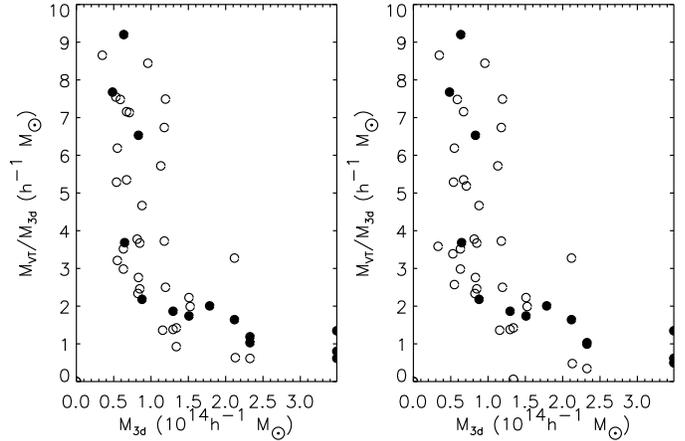}
\caption{\label{virial_fig} Comparison of the virial mass estimate
  $M_{\rm VT}$ to the true cluster mass $M_{\rm 3-D}$. Both panels of
  this figure display the ratio $M_{\rm VT}/M_{\rm 3-D}$ as a function
  of the true cluster mass $M_{\rm 3-D}$. In the left panel, the
  velocity dispersions before $3\sigma$ clipping are used to compute
  the virial-theorem based mass estimate, while in the right panel,
  the velocity dispersions after $3\sigma$ clipping are used. In both
  panels, the open circles refer to ${\cal R}=0$ clusters and the
  filled circles to ${\cal R}=1$ clusters.}
\end{figure}

Figure~\ref{virial_fig} displays the correlation of the virial mass
estimate with the true mass of the corresponding 3-D cluster. The left
and right panels show the ratio of the virial mass with the 3-D
cluster mass as a function of the 3-D mass before and after $3\sigma$
clipping, respectively.

\begin{table*}
\centering
\caption{\label{mass_estimates} Comparison of statistical parameters
  for the ratio $M_{\rm VT}/M_{\rm 3-D}$ and $M_{\zeta}/M_{\rm 3-D}$
  for different sub-samples of masses estimated via the velocity
  dispersion and the $\zeta$-statistics.}
\begin{tabular}{@{}lccc@{}}
  sample & mean & median & standard deviation \\ [10pt]
  complete optical sample before $3\sigma$ clipping &
  4.90 & 3.52 & 1.91 \\
  Abell cluster ${\cal R}=1$ before $3\sigma$ clipping &
  3.84 & 1.86 & 1.10 \\
  Abell cluster ${\cal R}=0$ before $3\sigma$ clipping &
  5.44 & 3.73 & 2.67 \\
  \\
  complete optical sample after $3\sigma$ clipping &
  4.53 & 3.39 & 2.01 \\
  Abell cluster ${\cal R}=1$ after $3\sigma$ clipping &
  3.58 & 1.86 & 1.18 \\
  Abell cluster ${\cal R}=0$ after $3\sigma$ clipping &
  5.01 & 3.67 & 2.80 \\
  \\
  complete lensing sample & 1.27 & 1.05 & 0.34 \\
  $S\ge5$                 & 1.23 & 1.13 & 0.34 \\
  $5\ge S\ge4$            & 1.32 & 1.02 & 0.31 \\
\end{tabular}
\end{table*}

The first thing to notice is that the masses of clusters with richness
class ${\cal R}>0$ are less severely overestimated than masses of
clusters with lower richness. This holds true for the mass estimates
before and after $3\sigma$-clipping for both the mean and the median,
as can be seen in Tab.~\ref{mass_estimates}. The second thing readily
seen in Fig.~\ref{virial_fig} and Tab.~\ref{mass_estimates} is the
large dispersion of the underlying distribution. This dispersion is
smaller for clusters with higher richness class than for clusters with
the lowest richness class ${\cal R}=0$ considered. We also note that
this dispersion is hardly affected by the $3\sigma$-clipping
procedure. The only effect of the clipping procedure is to reduce the
average of the estimated cluster masses irrespective of the richness
class. A third trend to be seen in Fig.~\ref{virial_fig} is that the
overestimation of the mass is generally more severe for 3-D clusters
with lower mass. For the most massive clusters in the sample
($M_{3-D}\la2.0\times 10^{14}\,h^{-1}\,M_\odot$), the continuation of
this trend in some cases leads to an {\em underestimation\/} of the
masses, as can be seen in the right-hand side of each panel in
Fig.~\ref{virial_fig}. The $3\sigma$ clipping procedure fails to
correct for the mass overestimates. When comparing
Fig.~\ref{virial_fig} and Tab.~\ref{mass_estimates} to Fig.~15 of Cen
(1997), one has to keep in mind the different selection procedure for
clusters or groups of galaxies in both studies, but on the whole the
results are consistent.

The behaviour described above can largely be attributed to the
influence of projection effects on the velocity dispersion. Generally,
the inclusion of field galaxies and unvirialised sub-clumps broadens
the distribution and leads to distributions which deviate
significantly from Gaussian shape, as illustrated by the examples in
Sect.~\ref{structure}. The clipping procedure is successful when the
amount of contamination along the line-of-sight is low or moderate,
but the algorithm fails to remove larger sub-clumps projected onto the
main cluster which can significantly broaden the distribution,
sometimes even making it bimodal. In some cases it is possible that
the clipping procedure removes galaxies belonging to the 3-D cluster,
thus contributing to an underestimation of the mass.

Even though we expect from the studies of Frenk et al.~(1990) and van
Haarlem et al.~(1997) that the high-velocity tail of the velocity
distribution is severely overestimated, the effect on the mass
estimate is most pronounced for galaxy clusters with lower mass. This
is due to the fact that they are more easily overestimated with
respect to their true dispersion.

\begin{figure}[ht]
  \includegraphics[width=\columnwidth]{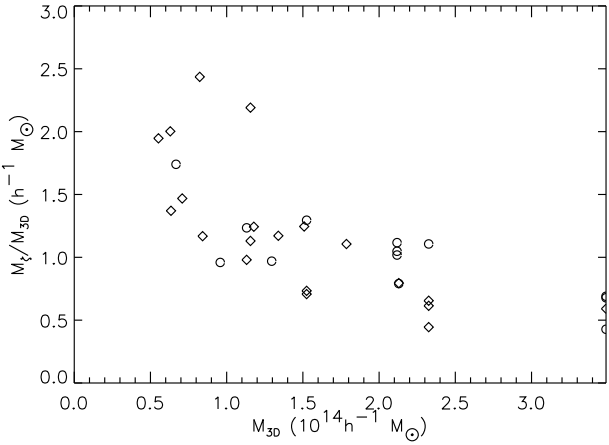}
\caption{\label{zeta_fig} Comparison of the $\zeta$-statistics mass
  estimate to the true cluster mass $M_{\rm 3-D}$. Diamonds and
  circles refer to clusters identified with a $S\ge5$ and $S\ge4$,
  respectively. Clusters detected at lower $S$ are excluded because of
  their high contamination.}
\end{figure}

The $\zeta$-statistics as compared to Abell's criterion leads to
smaller overestimates of the 3-D cluster masses as shown in
Fig.~\ref{zeta_fig}. Interpreting the quantitative results of the
$\zeta$-statistics mass estimate, one has to keep in mind two
competing effects: On the one hand, the $\zeta$-statistics, like every
gravitational-lensing based method, measures all the mass along the
line-of-sight to the cluster; on the other hand, it gives a lower
bound to the cluster mass. In combination, these two competing effects
lead to fairly moderate mass overestimates, as can be seen in
Tab.~\ref{mass_estimates}. This also explains the difference to the
lensing mass estimates given in the paper by Cen (1997). There, all
masses along the line of sight are added up under the assumption of a
perfect lensing reconstruction method with an otherwise calibrated
mass-sheet degeneracy.

The other interesting feature in Fig.~\ref{zeta_fig} and
Tab.~\ref{mass_estimates} is the low dispersion of the underlying
distribution. This dispersion does not depend sensitively on the
$S$-value at which the clusters are detected. (Clusters detected with
$S<4$ where excluded here because of their large contamination.) The
dispersion is typically less than a third of the dispersion in the
Abell samples.

As for clusters detected with Abell's criterion, masses of small 3-D
clusters are more strongly overestimated than for more massive 3-D
clusters. This is due to the fact that the proportion of contaminating
matter to the 3-D cluster mass is higher for less massive 3-D objects
than for the extremely massive objects. For the intermediate-mass
objects, the fact that the $\zeta$-statistics only gives lower bounds
to the masses partially outweighs this effect. There, the masses are
even slightly underestimated.

\begin{figure}[ht]
  \includegraphics[width=\columnwidth]{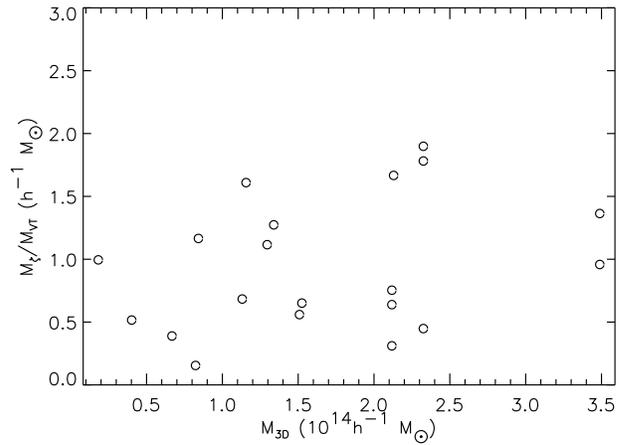}
\caption{\label{vir_zeta_fig} Comparison of the $\zeta$-statistics
  mass estimate to the virial mass estimate $M_{\rm VT}$. Only such
  clusters are included which are detected with both methods, Abell's
  criterion and $S$-statistics. Only clusters with $S>4$ are
  considered.}
\end{figure}

Investigating the relationship between velocity-based mass estimates
and the gravitational lensing based $\zeta$-statistics in
Fig.~\ref{vir_zeta_fig}, we see that the $\zeta$-statistics gives on
average smaller estimates of the 3-D cluster masses than the virial
theorem. Again we stress that this is due to the fact that the
$\zeta$-statistics is derived under the assumption of an empty outer
annulus, restricting it to estimate lower bounds to the masses. The
dispersion between the ratio of $\zeta$-statistics mass to virial mass
is large, which is due to the large underlying dispersion in the
virial mass estimate.

\section{Conclusions}

We investigated for the first time with simulated data whether
mass-selected galaxy cluster samples are more reliable than samples
constructed via Abell's criterion. Selection of clusters by mass is
possible through their gravitational lensing effects, in particular
through the coherent image distortion patterns imposed on faint
galaxies in their background. As mentioned in the introduction, image
distortions trace the gravitational tidal field of a lens rather than
its mass, and it is in that sense that we speak of ``mass-selected''
cluster samples. The second-order aperture-mass statistics $M_{\rm
ap}$ was used which is particularly suitable for detecting dark-matter
haloes. The signal-to-noise ratio, $S$, of $M_{\rm ap}$ provides a
straightforward detection criterion. We also compared the performance
of cluster mass estimators based on cluster-galaxy kinematics and
gravitational lensing. Our results can be summarised as follows.

\begin{enumerate}

\item As already found in previous studies, Abell clusters are
  severely affected by projection effects. This not only concerns the
  selection of Abell clusters, but also mass estimates based on galaxy
  kinematics and the virial theorem, indicating that the velocity
  dispersion is also hampered by projection effects. A second reason
  for the failure is the fact that the assumption of dynamical
  equilibrium is not justified in at least some of the clusters. The
  projection effects are worse for clusters and groups of lower
  richness class.

\item Clusters detected with a high significance $S$ of $M_{\rm ap}$
  are less affected by projection effects than typical Abell-selected
  clusters. Like Abell cluster samples, the mass-selected cluster
  samples are generally incomplete: Samples of clusters detected above
  a certain $S$ threshold typically do not encompass all
  three-dimensional clusters present in the simulation; some clusters
  have lower $S$. However, the completeness of the samples can be
  increased by lowering the $S$ threshold. We therefore investigated
  the effect of varying the $S$ threshold on the samples. Completeness
  of $\approx100\%$ can be achieved for massive three-dimensional
  cluster samples ($M\ga10^{14}\,h^{-1}\,M_\odot$) by varying
  $S\ga4$. Then, the samples also contain a substantial fraction of
  spurious detections, most of which correspond to real clusters with
  smaller masses. Generally, there is a trade-off between completeness
  and the contamination by spurious detections. More complete cluster
  samples are more heavily contaminated by spurious clusters, and the
  balance can be adapted choosing the $S$ threshold. It should be
  noted that the exact thresholds on $S$ depend somewhat on the choice
  of the weight function entering the definition of $S$ (cf.~the
  discussion in Sect.~2).

\item While qualitatively the same trend is also observed in
  Abell-selected cluster samples, the $S$-statistics generally
  performs significantly better than Abell's criterion: Higher
  completeness can typically be achieved with a lower fraction of
  spurious detections. For instance, cluster samples detected at
  $S\ge4$ contain all of the most massive clusters in the simulation
  and 65\% spurious detections, while Abell samples with richness
  ${\cal R}\ge0$ encompass only about two-thirds of the most massive
  clusters and 82\% spurious detections.

\item Lensing-based mass estimates are significantly more accurate
  than mass estimates based on cluster-galaxy kinematics and the
  virial theorem. Virial masses are typically biased high because
  line-of-sight velocity distributions are broadened by projection
  effects. Lensing also adds up mass in front of and behind the
  clusters, but the bias is less severe. The standard deviation from
  the true (three-dimensional) mass of the lensing mass estimate is
  smaller by a factor of three or more than that of the virial mass
  estimates. It should, however, be noticed that the accuracy of
  lensing-based mass estimates depends on the depth of the
  background-galaxy sample and other observational effects. While the
  mass estimates based on the $\zeta$ statistics are accurate to
  within $\approx30\%$ in our simulations, they may well be less
  accurate under realistic observational conditions.

\end{enumerate}

Our study underestimates projection effects because of the limited
size of the simulation volume. This affects both the optical and the
lensing-based cluster selection. Yet it appears that selection of
clusters by mass yields more reliable cluster samples than optical
cluster selection, and, more importantly, the quality of the samples
can be controlled by an objective criterion, namely the
signal-to-noise threshold imposed.

Selection of clusters by their gravitational-lensing effects is
comparable to selection by their X-ray luminosity. In essence, both
methods measure weighted projections of the Newtonian cluster
potential along the line-of-sight. However, lensing-based cluster
detections only require sufficiently deep imaging of wide fields in
optical or near-infrared wave bands, and detection algorithms can then
be applied in a straightforward manner. In particular, lensing can
detect clusters out to significantly higher redshifts than X-ray
surveys. What is more, lensing-based mass estimates do not rely on any
assumptions on the composition and physical state of the cluster
matter, in contrast to X-ray mass estimates. It can therefore be
expected that reliable, mass-selected cluster samples at moderate to
high redshifts can be constructed in the near future from upcoming
deep, wide-field surveys with a straightforward, well-controlled
algorithm, and that the accuracy of cluster mass estimates will
generally be substantially improved.

Lensing-based cluster detection algorithms like that based on the
$S$-statistics can be utilised for cosmology without reference to
actual 3-D clusters. Instead, one could define a ``cluster''
operationally as something visible as a sufficiently high peak in an
$S$ map, and then compare model predictions with observations
(cf.~Kruse \& Schneider 1999). If, however, emphasis is laid on
clusters as individual entities, it needs to be clarified how well
different selection criteria fare in detecting true clusters. Our
study has shown that selection of clusters by means of gravitational
lensing techniques can be adapted such that the resulting samples are
superior to Abell-selected samples in terms of completeness, spurious
detections, and the quality of mass estimates.

\section*{Acknowledgments}

We thank the GIF Consortium for making their $N$-body simulation
available to us, and Peter Schneider and an anonymous referee for
their many helpful comments and valuable suggestions.

\appendix

\section{Structure of further representative clusters}

We give a few more examples of line-of-sight structures of
$S$-selected galaxy clusters here.

\subsection{$S$-statistics: $S\ge5$}

\begin{figure}[ht]
  \includegraphics[width=\columnwidth]{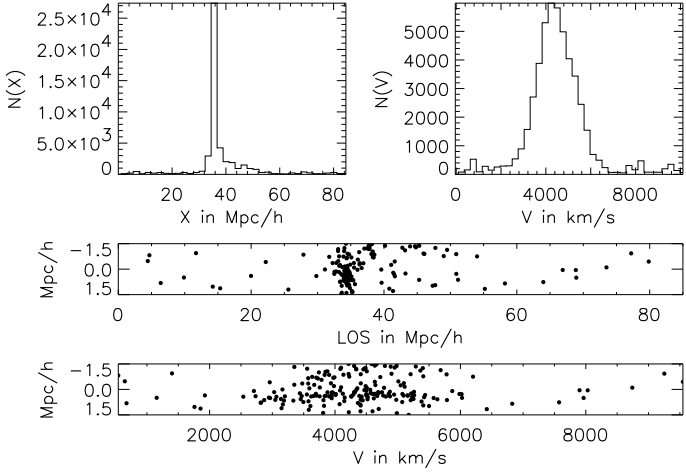}
\caption{\label{cluster_25x} Structure of a massive 3-D cluster
  detected with $S=9.6$. See the caption of Fig.~\ref{cluster_8x} for
  a description of the panels.}
\end{figure}

A second example for a cluster with high $S$ is given in
Fig.~\ref{cluster_25x}. The cluster is detected at $S=9.6$. The
particle distribution in real space is broad and dominated by a
massive 3-D cluster with a mass of $2.1\times
10^{14}\,h^{-1}\,M_\odot$. This cluster is detected as an Abell
cluster in projection, but the main 3-D cluster by itself already
passes the luminosity threshold of a 3-D Abell cluster. In contrast to
the first example, the velocity dispersion is hardly affected by
projection. The 3-D cluster has a velocity dispersion of $\sigma_{\rm
3-D}=884\,{\rm km\,s^{-1}}$, while the dispersion of the projected
cluster is $\sigma_{\rm 2-D}=779\,{\rm km\,s^{-1}}$. The higher-order
moments indicate a velocity distribution close to Gaussian shape for
both the 3-D cluster (${\cal S}_{\rm 3-D}=0.01$, ${\cal K}_{\rm
3-D}=-0.04$) and the projected cluster (${\cal S}_{\rm 2-D}=0.09$,
${\cal K}_{\rm 2-D}=-0.37$). All this reveals a fairly relaxed cluster
with low contamination.

\begin{figure}[ht]
  \includegraphics[width=\columnwidth]{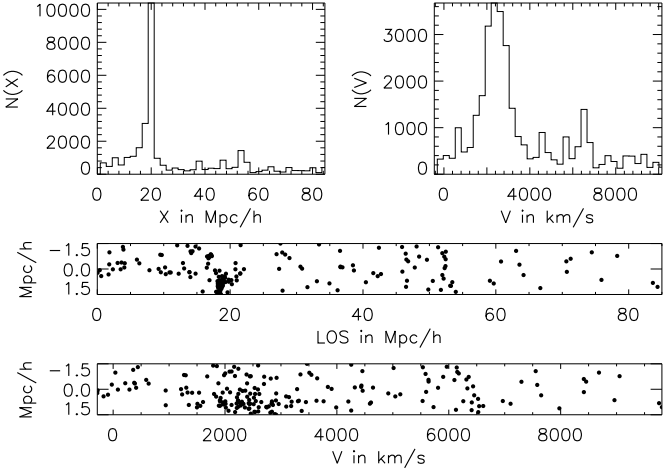}
\caption{\label{cluster_26y} Structure of a less massive 3-D cluster
  whose size is increased in projection due to matter concentrations
  along the line-of-sight. See the caption of Fig.~\ref{cluster_8x}
  for a description of the panels.}
\end{figure}

Almost all other clusters in this class show similar position and
velocity histograms. The only exceptions are the 2-D clusters
corresponding to less massive 3-D clusters. For one of these clusters
with relatively high $S=7.9$, the structure is given in
Fig.~\ref{cluster_26y}. Even though the position histogram is
dominated by a 3-D cluster, the distribution for this cluster is
broad, and there is a large amount of intervening matter with at least
four smaller clumps with masses of order
$M=10^{13}\,h^{-1}\,M_\odot$. Qualitatively, the {\em los\/} velocity
histogram looks artificially broadened by these clumps, and in fact
the velocity dispersion ($\sigma_{\rm 3-D}=651\,{\rm km\,s^{-1}}$) is
significantly increased in projection ($\sigma_{\rm 2-D}=860\,{\rm
km\,s^{-1}}$). The higher-order moments are also strongly affected by
this intervening matter (${\cal S}_{\rm 3-D}=-0.05$ and ${\cal K}_{\rm
3-D}=-0.64$ compared to ${\cal S}_{\rm 2-D}=1.19$ and ${\cal K}_{\rm
2-D}=0.87$). This cluster is detected as ${\cal R}=1$ Abell cluster
although it corresponds only to a ${\cal R}=0$ cluster in 3-D.

\subsection{$S$-statistics: $5>S\ge4$}

\begin{figure}[ht]
  \includegraphics[width=\columnwidth]{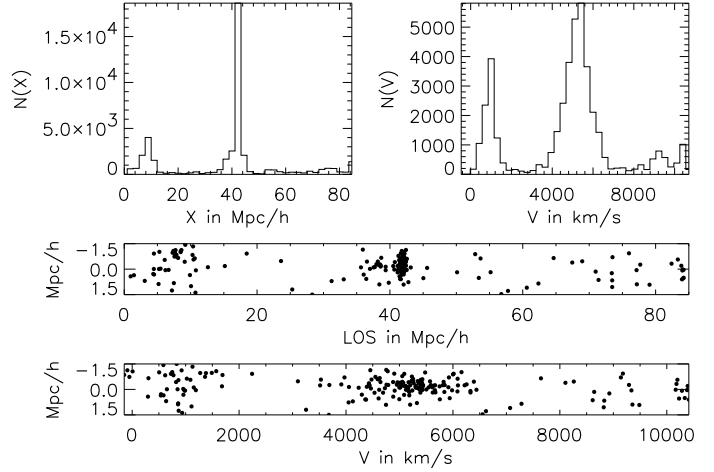}
\caption{\label{cluster_14x} Structure of a moderately contaminated
  3-D cluster with a $S=4.7$. See the caption of Fig.~\ref{cluster_8x}
  for a description of the panels.}
\end{figure}

Figure~\ref{cluster_14x} shows a cluster with $S=4.7$. It is
apparently only mildly contaminated by a clump $30\,h^{-1}\,$Mpc from
the main clump, which is a high-mass object with $M_{\rm
3-D}=2.3\times 10^{14}\,h^{-1}\,M_\odot$. The projected velocity
dispersion is almost unaffected ($\sigma_{\rm 2-D}=650\,{\rm
km\,s^{-1}}$ compared to $\sigma_{\rm 3-D}=635\,{\rm km\,s^{-1}}$),
and shows a bimodal feature, which is also reflected by the curtosis
of the projected cluster, ${\cal K}_{\rm 2-D}=0.51$, while the
velocity distribution of the 3-D cluster has a negative curtosis of
${\cal K}_{\rm 3-D}=-0.22$. Similarly, the skewness changes from
${\cal S}_{\rm 2-D}=-0.44$ to ${\cal S}_{\rm 3-D}=-0.03$. The cluster
is detected as a 2-D Abell cluster with richness class ${\cal R}=1$,
while the richness class of the 3-D cluster is ${\cal
R}=0$. Therefore, the richness class is inflated due to
projection. Even though this cluster shows some projection effects,
the corresponding 3-D cluster is massive and therefore clusters like
that should be included in a mass-limited sample.

\begin{figure}[ht]
  \includegraphics[width=\columnwidth]{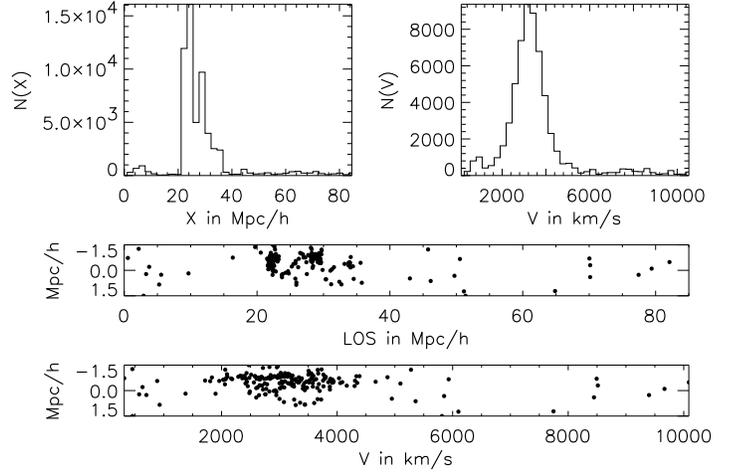}
\caption{\label{cluster_6z} Structure of a moderately contaminated 3-D
  cluster with $S=4.4$. See the caption of Fig.~\ref{cluster_8x} for a
  description of the panels.}
\end{figure}

The last example for this class is shown in
Fig.~\ref{cluster_6z}. Here, the $S$-map has a peak with $S=4.4$. The
position histogram shows a very broad peak with a secondary maximum on
top of the main peak. The corresponding 3-D cluster has a high mass,
$M_{\rm 3-D}=3.4\times 10^{14}\,h^{-1}\,M_\odot$. The projected
velocity distribution is only moderately skewed with ${\cal S}_{\rm
2-D}=0.22$ compared to the skewness of the main cluster alone, ${\cal
S}_{\rm 3-D}=0.17$. However, the curtosis of the projected peak,
${\cal K}_{\rm 2-D}=0.38$, even changes sign when compared to the 3-D
cluster, ${\cal K}_{\rm 3-D}=-0.27$. This cluster satisfies Abell's
criterion in projection, but the main peak has a lower richness class,
${\cal R}=0$.

\subsection{$S$-statistics: $4>S\ge3$}

\begin{figure}[ht]
  \includegraphics[width=\columnwidth]{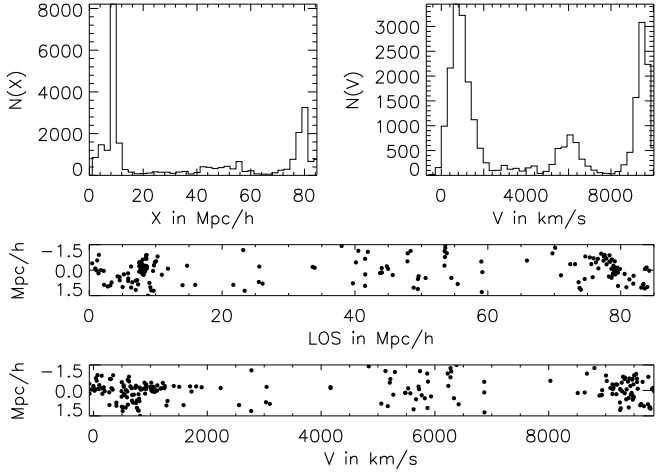}
\caption{\label{cluster_12z} Structure of moderately large 3-D group
  with $S$ between 3 and 4. The 3-D object is contaminated by
  projection along the line-of-sight, leading to an increased $S$ of
  $3.4$. See the caption of Fig.~\ref{cluster_8x} for a description of
  the panels.}
\end{figure}

Another example for a low-$S$ cluster detected at $S=3.4$ is displayed
in Fig.~\ref{cluster_12z}. This detection also corresponds to a 3-D
cluster with $M_{\rm 3-D}=6.6\times 10^{13}\,h^{-1}\,M_\odot$. Again,
the velocity distribution of this cluster is largely altered by the
considerable amount of intervening matter. The velocity dispersion
itself is inflated from $\sigma_{\rm 3-D}=504\,{\rm km\,s^{-1}}$ to
$\sigma_{\rm 2-D}=1134\,{\rm km\,s^{-1}}$. This is reflected by the
curtosis, which changes from ${\cal K}_{\rm 3-D}=-0.02$ to ${\cal
K}_{\rm 2-D}=-0.5$, while the skewness changes from ${\cal S}_{\rm
3-D}=0.18$ to ${\cal S}_{\rm 2-D}=-0.94$. Both low-$S$ examples are
neither 2-D Abell clusters nor do they pass the selection criteria for
Abell clusters in 3-D.

\begin{figure}[ht]
  \includegraphics[width=\columnwidth]{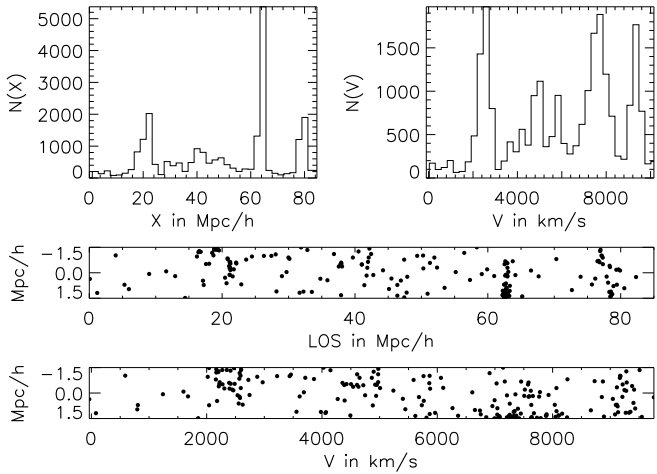}
\caption{\label{cluster_39x} Structure of a spuriously detected object
  which does not correspond to a 3-D cluster. The high contamination
  along the line-of-sight leads to $S=3.8$. See the caption of
  Fig.~\ref{cluster_8x} for a description of the panels.}
\end{figure}

The last example in Fig.~\ref{cluster_39x} with $S=3.75$ does not
correspond to a 3-D cluster or group with mass exceeding $M_{\rm
3-D}=10^{13}\,h^{-1}\,M_\odot$. Instead, one sees a large amount of
contaminating matter and smaller sub-clumps. This material is
responsible for the signal in the $S$ map. The velocity distribution
is characterised by three peaks with dispersion $\sigma_{\rm
2-D}=1235\,{\rm km\,s^{-1}}$, skewness ${\cal S}_{\rm 2-D}=-0.30$, and
curtosis ${\cal K}_{\rm 2-D}=-1.39$. Obviously, the contamination
along the line-of-sight is large enough to lead to the detection of an
Abell cluster with richness class ${\cal R}=0$.


\begin{thebibliography}{99}
\bibitem{ref:1} Abell G.O., 1958, ApJS, 3, 211
\bibitem{ref:2} Bartelmann, M., 1995, A\&A, 303, 643
\bibitem{ref:3} Bartelmann M., Narayan R., 1995, ApJ, 451, 60
\bibitem{ref:4} Bartelmann M., Steinmetz, M., 1996, MNRAS, 283, 431
\bibitem{ref:5} Bartelmann M., Narayan, R., Seitz, S., Schneider, P.,
  1996, ApJ, 464, L115
\bibitem{ref:6} Binney J., Tremaine S., 1987, Galactic Dynamics,
  Princeton University Press
\bibitem{ref:7} Bond J.R., Efstathiou G., 1984, ApJ, 285, L45
\bibitem{ref:8} Cen R., 1997, ApJ, 485, 39
\bibitem{ref:9} Cole S.M., Lacey C., 1996, MNRAS, 281, 716
\bibitem{ref:10} Couchman H.M.P., Thomas P.A., Pearce F.R., 1995, ApJ,
  452, 797
\bibitem{ref:11} Couchman H.M.P., Thomas P.A., Pearce F.R., 1996,
  astro-ph/9603116
\bibitem{ref:12} Davis M., Efstathiou G.P., Frenk C.S., White S.D.M.,
  1985, ApJ, 292, 371
\bibitem{ref:13} Fahlman G., Kaiser N., Squires G., Woods D., 1994,
  ApJ, 437, 56
\bibitem{ref:14} Frenk C.S., White S.D.M., Efstathiou G., Davis M.,
  1990, ApJ, 351, 10
\bibitem{ref:15} Gunn J., Gott J.R., 1972, ApJ, 176, 1
\bibitem{ref:16} Jenkins A., Frenk, C.S., Pearce, F.R., Thomas, P.A.,
  et al., 1998, ApJ, 499, 20
\bibitem{ref:17} Kaiser N., 1995, ApJ, 439, L1
\bibitem{ref:18} Kaiser, N., Squires, G., 1993, ApJ, 404, 441
\bibitem{ref:19} Kaiser, N., Squires, G., Broadhurst, T., 1995, ApJ,
  449, 460
\bibitem{ref:20} Kaiser, N., Wilson, G., Luppino, G., Kofman, L.,
  Gioia, I., Metzger, M., Dahle, H., 1998, ApJ submitted, preprint
  astro-ph/9809268
\bibitem{ref:21} Kauffmann, G.A.M., Colberg, J.M., Diaferio, A.,
  White, S.D.M., MNRAS submitted, preprint astro-ph/9805283
\bibitem{ref:39} Kruse, G., Schneider, P., 1999, MNRAS, 302, 821
\bibitem{ref:22} Lucey J.R., 1983, MNRAS, 204, 33
\bibitem{ref:23} Marzke R.O., Huchra J.P., Geller M.J., 1994, ApJ,
  428, 43
\bibitem{ref:24} Sarazin C.L., 1986, Rev. Mod. Phys., 58, 1
\bibitem{ref:25} Schechter P., 1976, ApJ, 203, 297
\bibitem{ref:26} Schneider P., 1996, MNRAS, 283, 837
\bibitem{ref:27} Schneider P., Kneib J.P., in: {\em The Next
  Generation Space Telescope: Science Drivers and Technological
  Challenges\/}, 34th Li\`ege Astrophysics Colloquium, June 1998,
  p.~89
\bibitem{ref:28} Schneider P., Seitz C., 1995, A\&A, 294, 411
\bibitem{ref:29} Schneider P., Ehlers J., Falco E.E., 1992,
  Gravitational Lenses, Springer-Verlag
\bibitem{ref:30} Schramm T., Kayser R., 1995, A\&A, 299, 1
\bibitem{ref:31} Seitz C., Schneider P., 1995, A\&A, 297, 287
\bibitem{ref:32} Seitz C., Schneider P., 1997, A\&A, 318, 687
\bibitem{ref:33} Seitz, S., Schneider, P., 1996, A\&A, 305, 383
\bibitem{ref:34} Squires, G., Kaiser, N., 1996, ApJ, 473, 65
\bibitem{ref:36} Swarztrauber P., 1984, in Golub G., ed., Studies in
  Numerical Analysis, 24, p.319
\bibitem{ref:35} van Haarlem M.P., Frenk C.S., White S.D.M., 1997,
  MNRAS, 287, 817
\bibitem{ref:37} White S.D.M., Efstathiou G., Frenk C.S., 1993, MNRAS,
  262, 1023
\bibitem{ref:38} Yahil A., Vidal N.V., 1977, ApJ, 214, 347
\end{thebibliography}
\end{document}